# Galactic Neutral Hydrogen Structures Spectroscopy and Kinematics: Designing a Home Radio Telescope for 21 cm Emission

Phelps, J

## Abstract

This study presents the methodology for creating a cost-efficient radio astronomy telescope that can be used to detect 21 cm emissions (f21 = 1420.405 MHz) and determine the distribution and kinematics of neutral hydrogen specifically in the Milky Way. By measuring the Doppler shifts of the 21 cm emission, the velocities of hydrogen clouds relative to Earth can be determined. This enables the identification of these clouds' movements, their positions within the galaxy's spiral arms, and their roles in the overall rotational dynamics of the Milky Way. The setup is designed to be accessible to amateurs, enabling others to conduct similar projects. The measurement apparatus consists of a 1-meter parabolic dish, a H1-LNA for 21 cm emissions, an SDR, and a Raspberry Pi. This paper also provides an overview of the data processing required to detect the hydrogen line and generate velocity profiles. Additionally, it examines RFI mitigation techniques, such as spectral filtering and instrument shielding, which enhance observational clarity even in urban environments like Los Angeles. This study also analyzes the observed velocities of different galactic arms, as well as measurements across the sky.

## Scientific Background:

The 21 cm atomic hydrogen (HI) line is a key spectral line in radio astronomy. Emission results from the hyperfine splitting of neutral hydrogen atoms. This occurs due to the interaction between the electron's spin and the proton's spin within the hydrogen atom. Spin refers to the intrinsic angular momentum of subatomic particles, such as electrons and protons. In a hydrogen atom, when the spins of the electron and proton are aligned (parallel), the system is in a slightly higher energy state. When the spins are anti-aligned (anti-parallel), the system is in a lower energy state. Hyperfine splitting occurs because of the small energy difference between these two states. When the hydrogen atom transitions from the higher energy state to the lower one, it emits a photon with a wavelength of 21 cm, corresponding to a frequency of 1420.405 MHz.

The 21 cm line is particularly valuable because neutral hydrogen is the most abundant element in the universe, making up much of the interstellar medium (ISM) in galaxies and ≥70% of normal matter in the universe. Unlike light in the optical spectrum, 21 cm radiation can penetrate dust and gas. This allows for the observation of regions otherwise hidden from view. This property allows astronomers to map the structure of galaxies, such as the Milky Way, and study the distribution and dynamics of neutral hydrogen clouds.

According to the Lambda-CDM (ΛCDM) cosmology model, dark matter is crucial for galaxy formation and structure, while dark energy drives the accelerated expansion of the universe. Astronomers can determine the distribution of hydrogen in galaxies through the 21 cm line, while also learning about the interplay between baryonic matter (such as hydrogen) and dark matter. The rotational dynamics of galaxies, derived from the velocities of hydrogen clouds, often suggest the presence of large amounts of dark matter, particularly in the outer regions where visible matter is sparse.

HI maps can be created using observations of the 21 cm line and can help shed light on the expansion history of the universe. By mapping the distribution of neutral hydrogen across different galaxies, large-scale structure of the cosmos can be traced. (Furlanetto, 2006). This form of mapping reveals patterns of galaxy clustering and movement over time, which can help refine models of cosmic expansion and the influence of dark energy on the acceleration of the universe's growth. Intensity mapping, like that carried out in the CHIME experiment, measures the collective brightness of hydrogen emission across large volumes of space without resolving individual galaxies. This method allows scientists to efficiently probe vast regions, capturing information about the distribution and density of hydrogen gas over cosmic time. (e.g., CHIME Collaboration 2022; Battye, Davies & Weller 2004; Wyithe & Loeb 2008; Chang et al. 2008) By understanding where and how much hydrogen exists, astronomers can piece together the life cycle of galaxies and how they interact with their surroundings, leading to a more comprehensive picture of cosmic evolution.

Cross correlation techniques that combine 21 cm hydrogen line observations with other data sources, such as galaxy catalogs can enhance the detection of hydrogen structures by correlating regions where both absorption and emission are present. This has the capability to elucidate details about the reionization process and epoch of reionization (EoR). (Kadota et al., 2019). By analyzing both the absorption and emission of hydrogen, and synthesizing data from experiments like the Hydrogen Epoch of Reionization Array (HERA) and galaxy surveys like eBOSS, light can be shed on how the universe transitioned from a dark, neutral state to one filled with luminous galaxies. (DeBoer, 2016; La Plante, 2022; Wolz, 2021). This approach also reveals how hydrogen clustering patterns evolved over time and offers insights into the expansion history of the universe. Through cross-correlation techniques, a better understanding of cosmic structures and the influence of dark matter helps to piece together a more comprehensive picture of how galaxies formed and how the universe expanded from its early stages to the present day.

Low-cost, at-home 21 cm detectors, like the one described in Fung et al. (2023), provide an avenue for interested students and amateurs to make their own measurements of this critical signal in a Galactic context.

# Experiment Design

The setups consist of an antenna, intermediary signal conditioning devices, and a computer. The system is built on the capabilities of a software defined radio (SDR) which enables the computer to understand the input from the telescope. SDRs transform radio signals into data streams, which can then be accessed through a standard USB connection to a computer. This setup provides substantial flexibility by handling tasks such as signal processing and modulation via software rather than dedicated hardware components.

The dish is a 1 meter parabolic dish traditionally intended for satellite data reception. The dish has a center frequency of about 1.4 GHz. The dish elevation was varied and drift scans were conducted.

The data (analog signal) collected from the dish is sent through a Low Noise Amplifier (LNA) followed by a Bandpass Filter (BPF), and then a second LNA. The signal is then converted into a data stream that can be understood by the computer by the Software Defined Radio (SDR). The data then undergoes further data processing to refine and extrapolate meaning from the data.

The antenna and computer manifold were installed on the base of a custom radio tower about 35 feet above ground level. The computer, SDR and LNA were placed as close as possible to the antenna to minimize loss. The entire setup was grounded to earth ground.

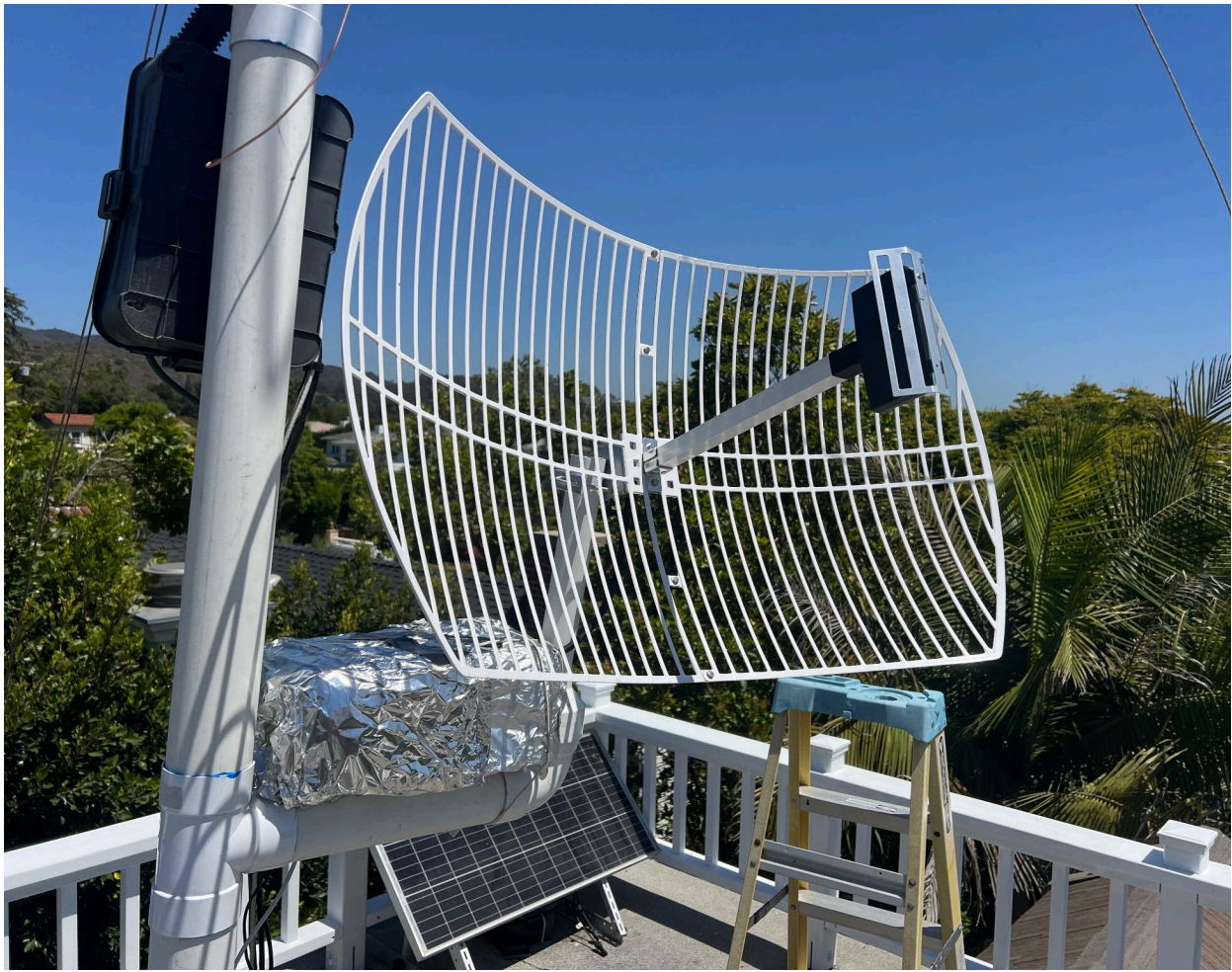

Figure 1a: 1 meter dish with tin foil shielding

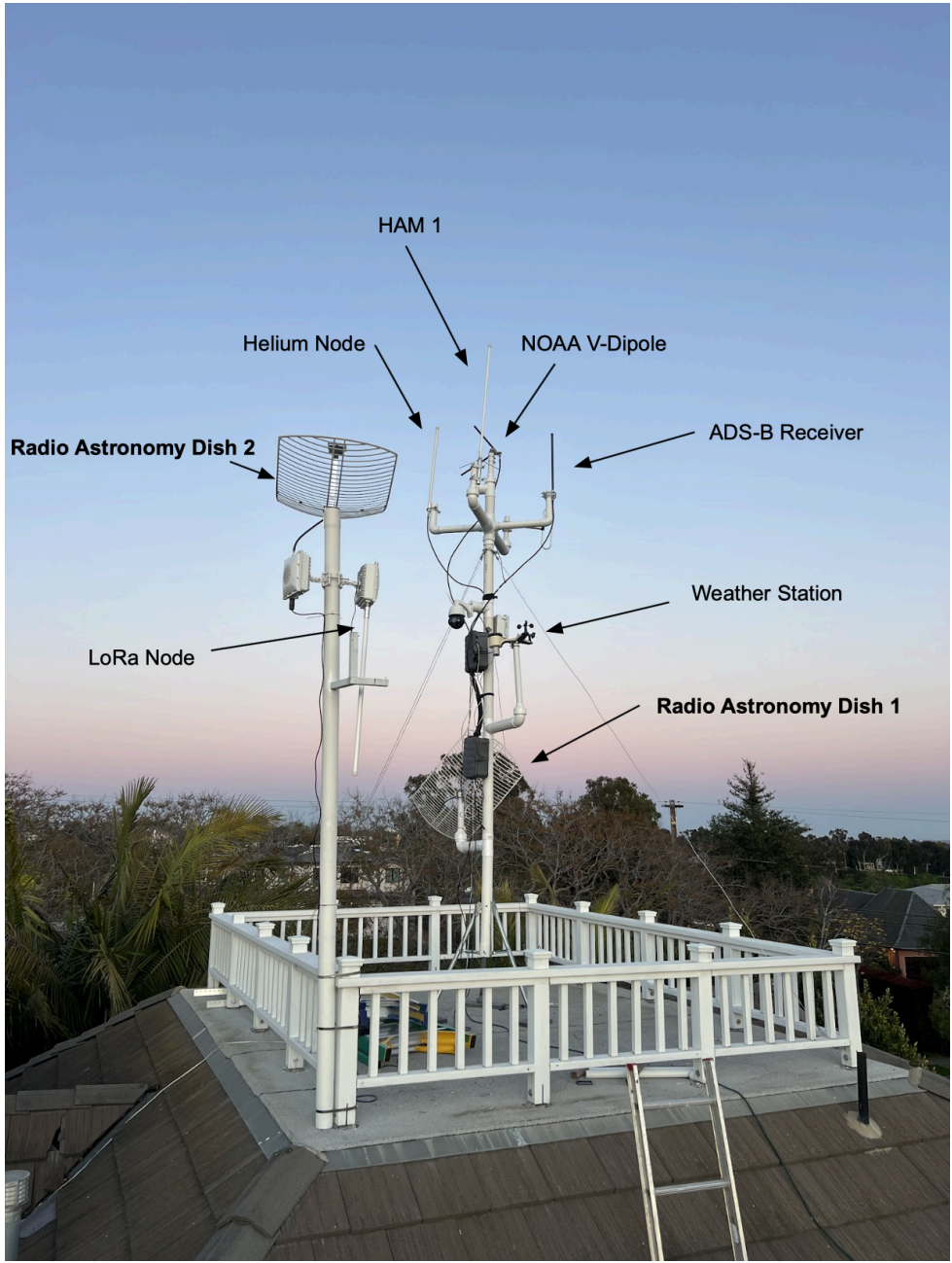

Figure 1b: The setup consists of 2 radio towers each with their own radio telescope. The primary tower yields Dish 1 which was primarily used for drift scans at multiple elevations facing due south. It is important to note that in this picture the dish is pointed and skewed for satellite data reception, not radio astronomy. Additionally, the shielded box that normally houses Dish 1's electronics is absent in this photo. Normally it would have no polarization skew. The telescope on the secondary tower is pointed 90 degrees vertical and was primarily used for dark sky, or cold observation/ calibration. Its electronics are housed in the shielded box directly below it. The usage of two dishes allows for verification of measurement and additional ease of use. All other antennas on the tower are labeled.

# Hardware:

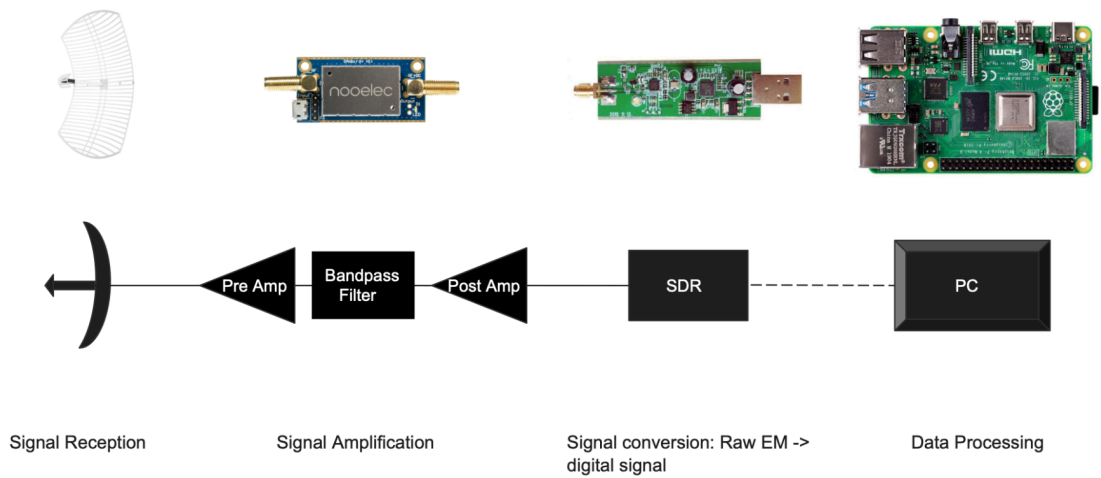

Figure 2: The Hardware pipeline consists of three parts: antenna, signal conditioners, and computer. The solid lines indicate LMR-400 cable (low loss microwave coax), whereas the dotted line represents USB 3.0. (Figure inspired by (Petitt, 2022)).

## Antenna

The observational setup consists of three parts. The antenna is a 1 meter parabolic dish. The dish is primarily used for satellite data reception and has a center frequency between 1300 and 1600 Mhz. This antenna is high gain and directive, which is good for more precise observations. However, the high gain means that the dish is susceptible to more noise than something like a horn antenna. The diameter of the dish (1 meter) and the frequency (1420.4 MHz) is known, so the estimated beamwidth of the antenna can be calculated using the equation below. The wavelength in meters of the target frequency 1420.4 MHz is about 0.2110619952 meters.

$$\text{BW} = \frac{70 \times \lambda}{D} \qquad \text{BW} = \frac{70 \times 0.211 \text{ m}}{1 \text{ m}} = \frac{14.7847 \text{ m}}{1 \text{ m}} = 14.78°$$

Figure 2a: The calculations show the beamwidth calculation formula where D is the diameter of the dish and λ is the wavelength of the signal.

The calculated beamwidth of the antenna is about 14.78°. Unlike horn antennas, parabolic dishes generally produce a more symmetrical, circular beam pattern. The use of a parabolic dish also means the antenna has sharp main lobes and short side lobes, allowing for higher directivity and gain.

Additionally, the stock antenna has about 1 foot of LMR-400 coaxial cable coming out of the feed. This introduces considerable noise. In an ideal setup the LNA would be connected directly to the feed and there would be no intermediate coaxial cable. The VSWR of the antenna was measured using a vector network analyzer (VNA) and is shown below.

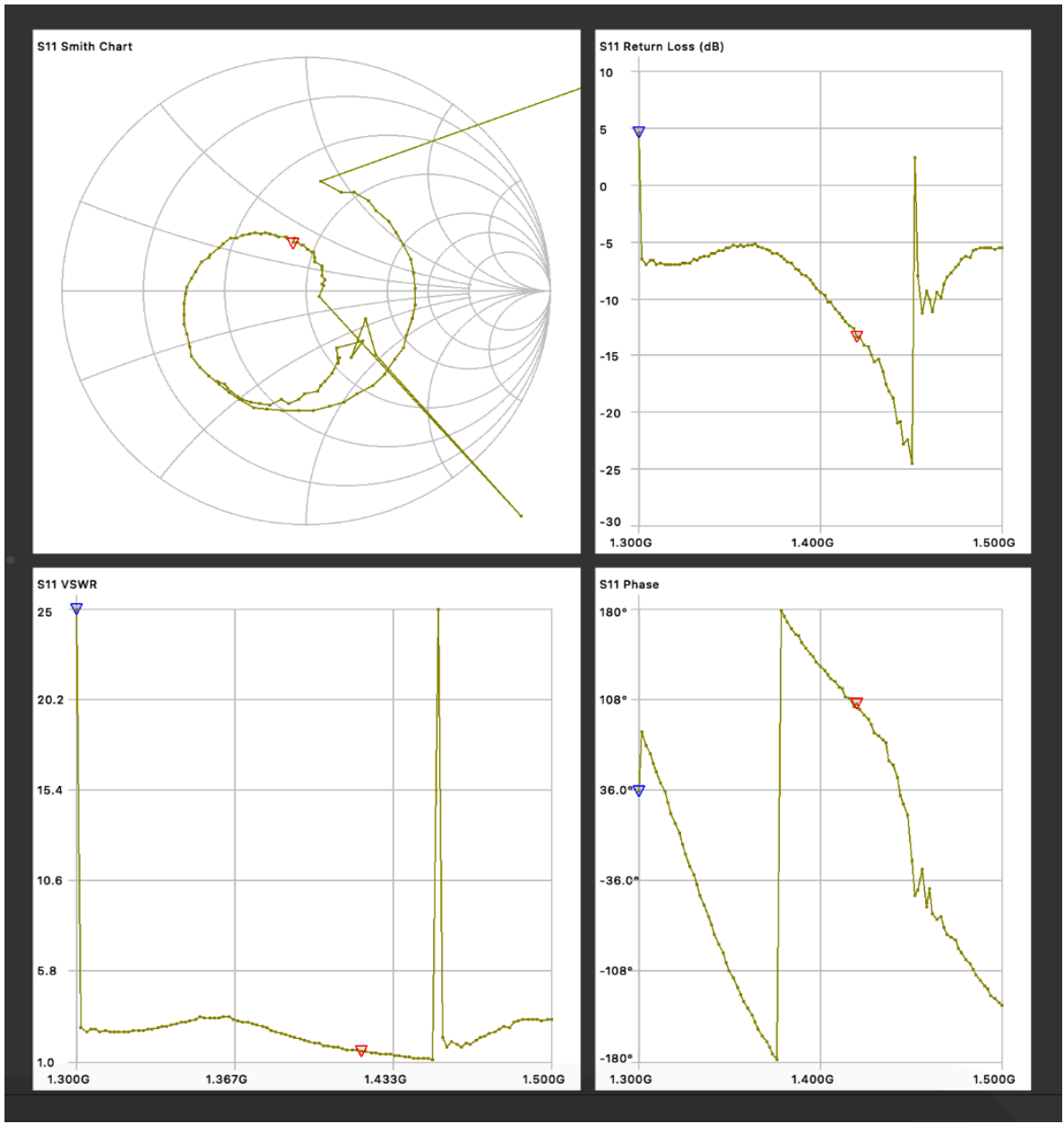

Figure 2b: A vector network analyzer (VNA) was used to test the antenna. The data shows impedance (Smith chart), VSWR, return loss, and phase.

| Frequency: | 1.42000 GHz |
|---|---|
| Impedance: | 41.5÷17.8 |
| **VSWR:** | **1.538** |
| **Return loss:** | **-13.475 dB** |
| S11 Phase: | 104.45° |
| Minimum VSWR: | 1.126 @ 1.45000GHz |
| Minimum Return loss: | -24.521 dB |

Figure 2c: The data values from the VNA.

The VNA data for the antenna illustrates a range 1.3-1.5 GHz. The Smith Chart indicates a generally good impedance matching near the target frequency, where the return loss is measured at -13.475 dB. This suggests a relatively low level of reflected power and efficient receiving energy transfer. The VSWR was measured at 1.538. This supports the antenna's good impedance matching, as it falls within the acceptable range (VSWR < 2). It can be seen that the antenna is optimized/has the lowest VSWR of 1.126 @ 1.45000GHz. The S11 phase plot shows a phase angle of 104.45°. This is consistent with the behavior of an antenna near resonance. The impedance at this frequency is 41.5+j17.8 Ω41.5+j17.8Ω which indicates a slightly inductive nature of the antenna. The antenna is relatively well-matched and performs effectively in the vicinity of the hydrogen line frequency (1.420 GHz).

## Signal Conditioners

The signal conditioning chain consists of a Noelec H1 low noise amplifier (LNA) to boost the signal at 1420.4 MHz and a RTL-SDR v2 software defined radio to convert the raw signal into digital signal the audio card within a computer can understand.

The Nooelec H1 LNA does +40dB of RF gain at 1420MHz and 65MHz with 3dB bandwidth. It has a relatively low noise figure of 0.8dB at 1420MHz. The LNA is also encased in an EMI shield to mitigate interference and environment noise. The LNA is also powered via a BIAS-T which means no external dc power connection is required and additional noise is circumvented. The H1 has two, isolated, shielded LNAs on either side of a bandpass filter.

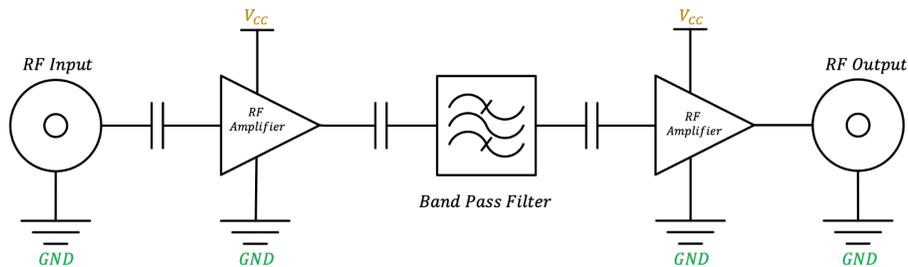

Figure 3a: Simplified Schematic from RTL-SDR data sheet showing components of Nooelec H1 LNA (*RTL-SDR Blog V3 Datasheet*.)

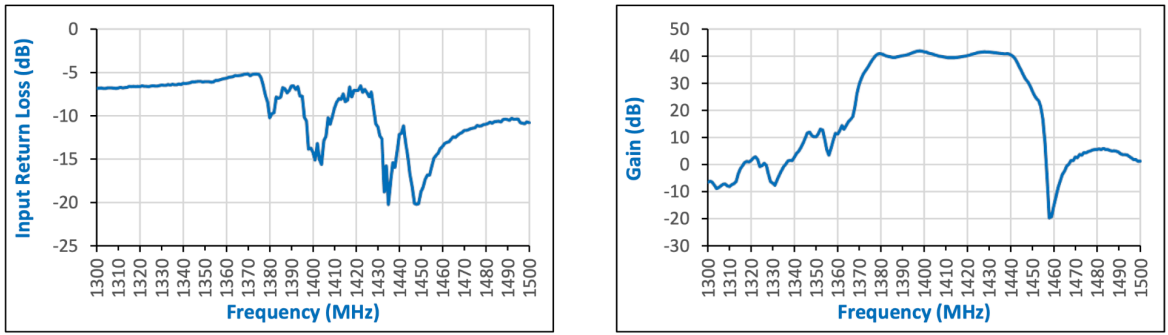

Figure 3b: Charts from H1 LNA datasheet showing the return loss and gain of the Nooelec H1 LNA (*SAWbird+ H1t Low Noise Amplifier (LNA) Ordering Information Description Simplified Schematic Features Applications*)

The RTL-SDR V2 works by converting the analog signal from the antenna into a digital signal that a computer can understand. This is accomplished with two integrated circuits (ICs): the RTL2832U and a tuner chip. The RTL2832U has two A/D converters and the tuner chip is an analog downconverter. Signals that pass through the tuner are then passed to the A/D converters. (*How RTL-SDR Dongles Work*, n.d.)

## Computer

A Raspberry Pi 4 with 8GB RAM and a 64-bit quad-core processor clocked at 1.5 GHz was utilized for the majority of the computational efforts. It was equipped with an air cooling radiator. The Pi was used for post conditioning data processing. The computer was powered using power over ethernet (POE) to further mitigate DC noise.

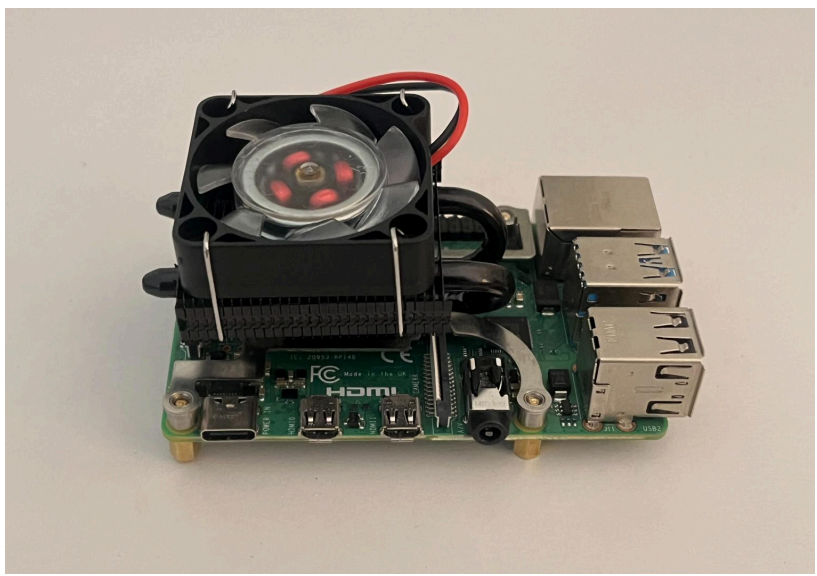

Figure 4: Raspberry pi 4 with cooler

# Software:

## OS and Primary Processing Software

The Raspberry Pi utilized a pre-made OS image created by Dr. Glenn Langston at the National Science Foundation. The OS contains multiple scripts for hydrogen line observation, calibration, and additional tools for data processing.

For primary data collection a python script combined with GNU radio was used. After passing through the SDR the data was processed by the following data processing pipeline. This includes the usage of fast fourier transforms (FFT) and vector medians. These transformations convert time-domain signals into their frequency components and help mitigate radio frequency interference (RFI).

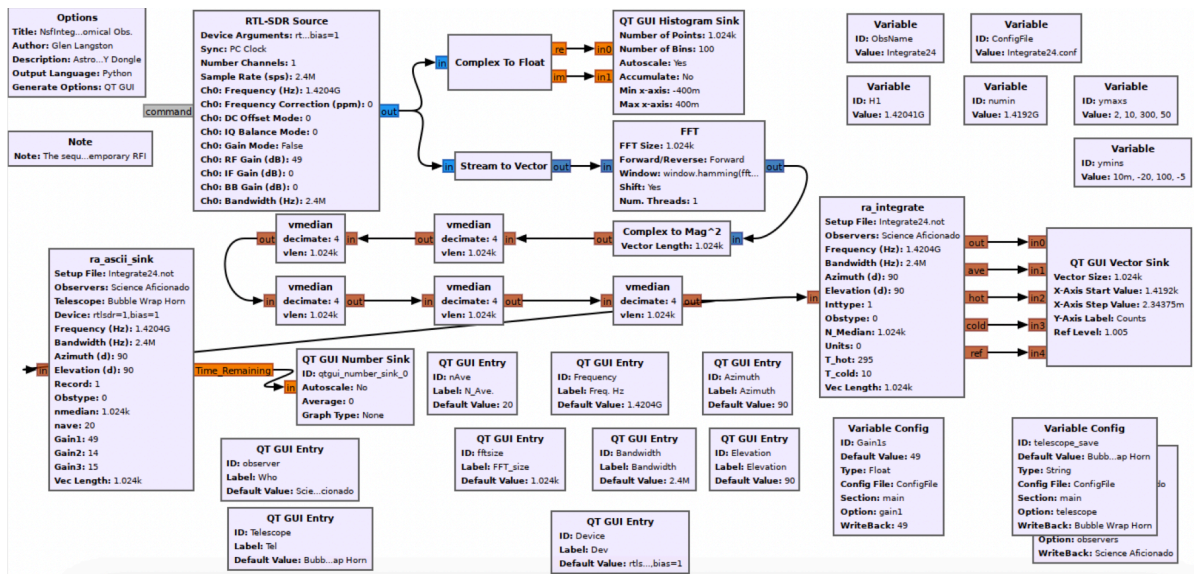

Figure 5a: This describes the layout of the NsfIntegrate design in GRC for use with the RTL SDR. The data flow starts from the OSMOSDR source at the top left, then passes through a block to create a complex vector and perform a Fourier Transform. The data rate is reduced using six Vector Median blocks, each of which takes four input vectors and produces one output vector. This process reduces the data rate from one new spectrum every 0.0003 seconds to one every 1.4 seconds, lowering the CPU load for plotting and averaging. This allows all data to be captured using a modest multi-core computer. The filtered data are then sent to the Ascii_Sink block for data writing and to the RAI block and plotter for real-time monitoring of the observations. (Langston & Bandura, 2018)

The Fast Fourier Transform (FFT) allows for a more efficient calculation of the Discrete Fourier Transform (DFT), which allows for the conversion from time domain to frequency domain. For hydrogen line emissions, the FFT is particularly useful because it converts the time-domain signal captured by the antenna into its frequency components while being more efficient than a DFT. The hydrogen line, which exists around 1420.4 MHz, can be identified as a specific peak in the frequency spectrum generated by the FFT. An FFT essentially speeds up a DFT by taking advantage of the periodicity within the DFT's calculations.

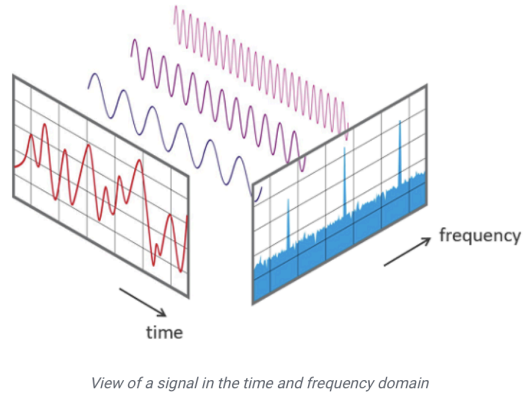

*View of a signal in the time and frequency domain*

Figure 5b: Visualization of time domain to frequency domain conversion
(*Fast Fourier Transformation FFT - Basics*)

The DFT is defined as a sequence x[n] of length N representing the time-domain signal received by the antenna.

$$X[k] = \sum_{n=0}^{N-1} x[n] \cdot e^{-j\frac{2\pi}{N}kn}$$

The FFT algorithm optimizes this calculation by splitting the sum into even and odd indexed parts

$$X[k] = \sum_{n=0}^{N/2-1} \left( x[2n] \cdot e^{-j\frac{2\pi}{N}2kn} + x[2n+1] \cdot e^{-j\frac{2\pi}{N}k(2n+1)} \right)$$

The equation can then be simplified

$$X[k] = \text{DFT}_{N/2}(x_{\text{even}}) + e^{-j\frac{2\pi k}{N}} \cdot \text{DFT}_{N/2}(x_{\text{odd}})$$

Where $x_{even}$ and $x_{odd}$ are the sequences of even and odd indexed elements, respectively. By recursively applying this process, the FFT allows for efficient identification of the hydrogen line frequency within the data, facilitating the analysis of signals captured by the antenna.

Another primary component of the signal processing pipeline is the employment of vector medians. Vector medians are used to determine the central tendency of a set of vectors in multi-dimensional space. For small sets of vectors, it is calculated by sorting the vectors dimension-wise and selecting the median value in each dimension, or by averaging the middle two values. For larger sets, an approximate vector median is determined by summing all vectors and then subtracting the vectors with the minimum and maximum magnitudes. This reduces the influence of outliers and unwanted RF noise, while providing an estimate of the central vector that best represents the overall data set. The vector median highlights the central tendency of the vectors while filtering out erratic noise. This leads to a more stable and accurate representation of the underlying signal and enhances the clarity and reliability of the processed data.

## Sky Visualization and Tracking

To verify that observations were accurate, and to visually verify what part of the sky the antenna was pointed at, Stellarium was used. This software is able to render an accurate depiction of the sky based on geographical coordinates and elevation. This tool also assisted with verifying that the cold or dark sky calibrations were done in a "dark" part of the sky.

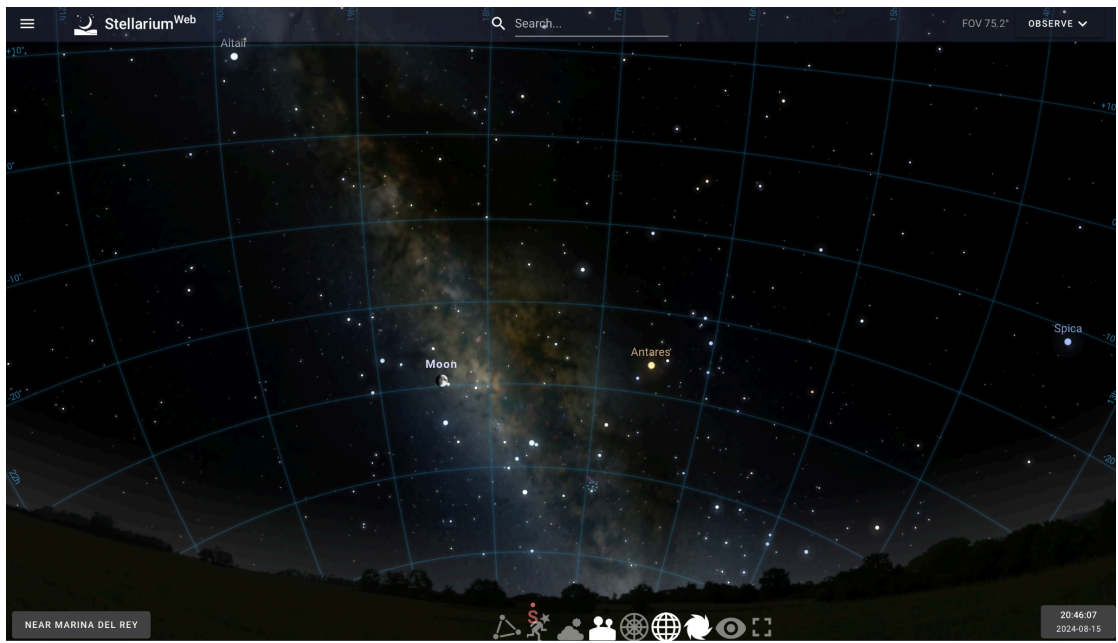

Figure 6: Stellarium generation of night sky. The tool uses geographical coordinates to generate a realistic replication of what the sky actually looks like and what objects are in sight. (*Stellarium*, n.d.)

# Observation:

## Shielding and RFI Mitigation

Environment noise and RFI decreases accuracy of measurement and negatively impact data. This leads to less differentiation between line profiles and background noise and decreases the SNR. RFI can be mitigated system side by encasing the electronics in grounded conductive material. Because the antennas were placed above a suburban house, there were many possible sources of RFI. To mitigate this the electronics were stored in shielded grounded boxes. This helped ensure that additional noise was dissipated into the ground and prevented from entering the electronics. In the case of this experiment, a plastic box was wrapped in 0.02 mm of tinfoil. This number was achieved by calculating the minimum thickness for signal penetration at the frequency. The foil not only protected against EMI, but also provided thermal insulation. This foil was grounded to earth ground.

1. Skin Depth Calculation

The skin depth ($\delta$) represents the distance over which the electromagnetic wave's intensity decreases to 1/e of its initial value when traveling through a conductive material. It is calculated using the formula

$$\delta = \sqrt{\frac{2}{\omega \mu \sigma}}$$

Where:

$\omega = 2\pi f$ is the angular frequency of the wave
$\mu = \mu_0 \mu_r$ is the magnetic permeability of the material
$\sigma$ is the electrical conductivity of the material

For aluminum:

- $f = 1420$ MHz (frequency of the hydrogen line)
- $\mu_0 = 4\pi \times 10^{-7}$ H/m (permeability of free space)
- $\mu_r = 1$ (relative permeability of aluminum)
- $\sigma = 3.5 \times 10^7$ S/m (conductivity of aluminum).

2. Angular Frequency ($\omega$)

The angular frequency is given by:

$$\omega = 2\pi f$$

3. Attenuation Constant (α)

The attenuation constant (α) is the reciprocal of the skin depth:

$$\alpha = \frac{1}{\delta}$$

4. Intensity Attenuation

The relative intensity (I) of the electromagnetic wave after passing through a material of thickness x is given by the exponential decay model:

$$I = I_0 e^{-\alpha x}$$

Where:

- $I_0$ is the initial intensity
- x is the thickness of the material

5. Parameters for Aluminum (Tinfoil)

Using the values for aluminum:

- $\delta = \sqrt{\frac{2}{(2\pi \times 1420 \times 10^6)(4\pi \times 10^{-7})(3.5 \times 10^7)}} \approx 6.9 \times 10^{-6}$ m (skin depth).

Therefore, the attenuation constant is:

$$\alpha = \frac{1}{\delta} \approx 1.45 \times 10^5 \text{ m}^{-1}$$

6. Final Equation for Intensity

Substituting α into the intensity equation:

$$I = I_0 e^{-1.45 \times 10^5 \times x}$$

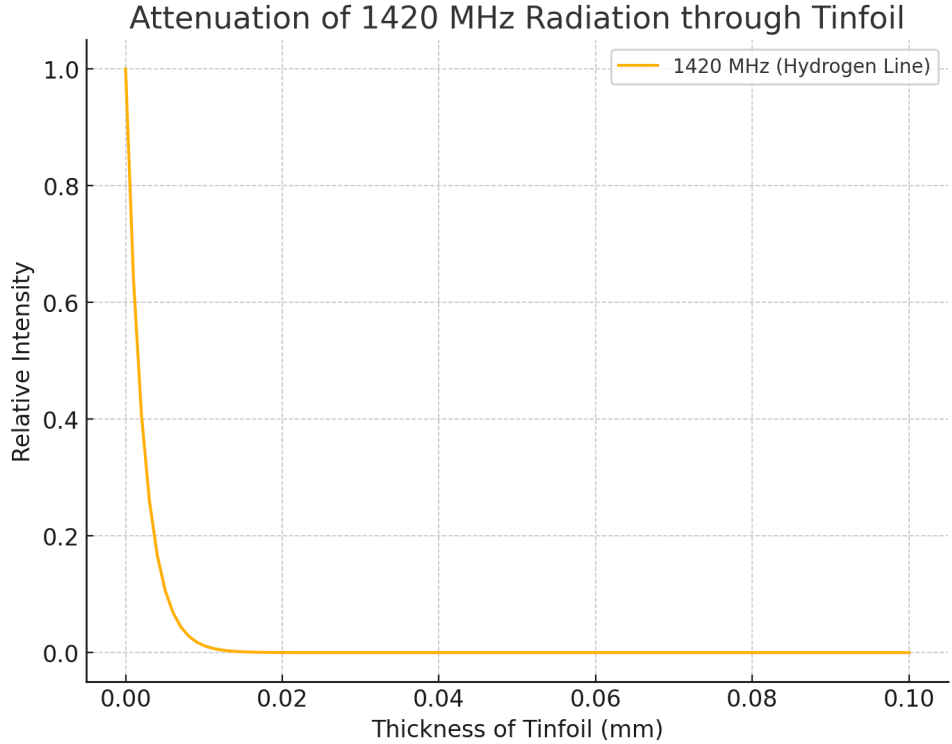

Figure 7: Attenuation curve

# Calibration

The calibration process was used to convert digital data, measured in arbitrary units called "counts," into physically meaningful quantities like temperature. The observed signal, $C_{Measured}$ (in counts), is related to the sky brightness temperature, $T_{Sky}$, and the system temperature, $T_{Sys}$, both measured in Kelvins. This relationship is given by the equation:

$$C_{measured} = S \cdot (T_{sys} + G \cdot T_{sky})$$

Here, S is the scale factor, which converts physical temperature into digital counts, and G is the gain factor. While this calibration was intended for usage with a horn antenna, it is still viable with a dish. Using a dish antenna instead of a feedhorn and waveguide probe means that G reflects the efficiency of the dish in focusing and collecting the signal. This gain factor depends on the dish's size, surface accuracy, and alignment. While the range of G remains from 0 to 1, a dish antenna generally has higher gain due to its larger collection area compared to a feedhorn.

Calibration involves measuring "hot" load counts $C_{Hot}$ and "cold" load counts $C_{Cold}$ using known temperatures (295 K for the ground and 15 K for the sky, respectively). These measurements allow for determining the system temperature $T_{Sys}$ by comparing the hot and cold load counts, calculated as:

$$T_{sys}/G = \frac{C_{Hot} \cdot (T_{Hot} - T_{Cold})}{\Delta C} - T_{Hot}$$

Where $\Delta C = C_{Hot} - C_{Cold}$

The sky brightness temperature at a specific location is then calculated by subtracting the system contribution from the observed counts, using the equation:

$$T_{sky}/G = \frac{C_{sky} \cdot (T_{Hot} - T_{Cold})}{\Delta C} - T_{Hot} - T_{sys}$$

Data reduction software, written in Python, processes the observation data by applying these equations, allowing the conversion of raw counts into temperatures, which results in accurate measurements of the sky's brightness. (Langston, 2015)

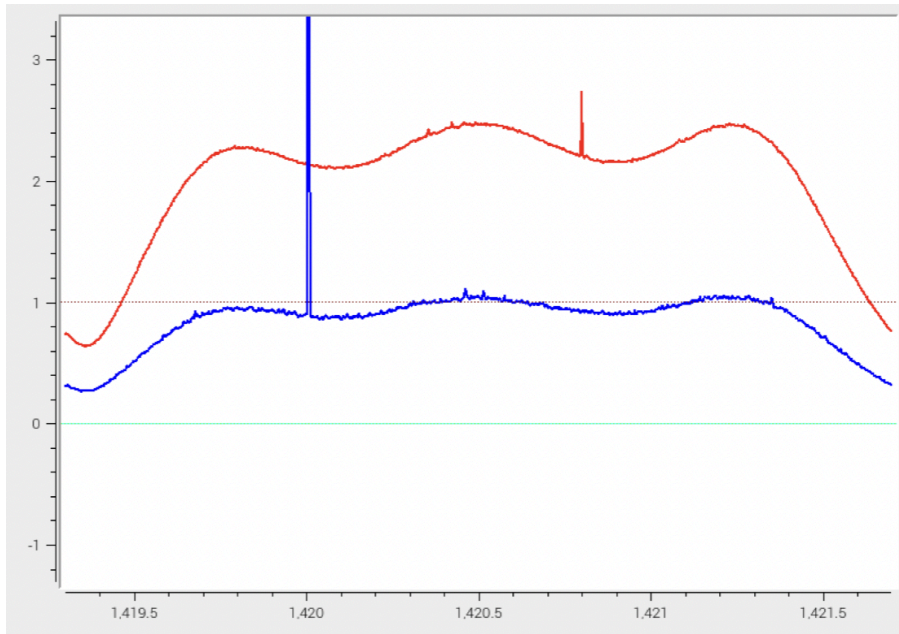

Figure 8: The hot (red line) and cold (blue) load observations used for calibration

# Initial Measurements

Initial observations and calibration were carried out using a software program within the OS called NsfIntegrate24. NsfIntegrate24 was mainly used as the data collection mechanism, but it contains a GUI and multiple plots which can be used to verify reception of the hydrogen line at the target frequency.

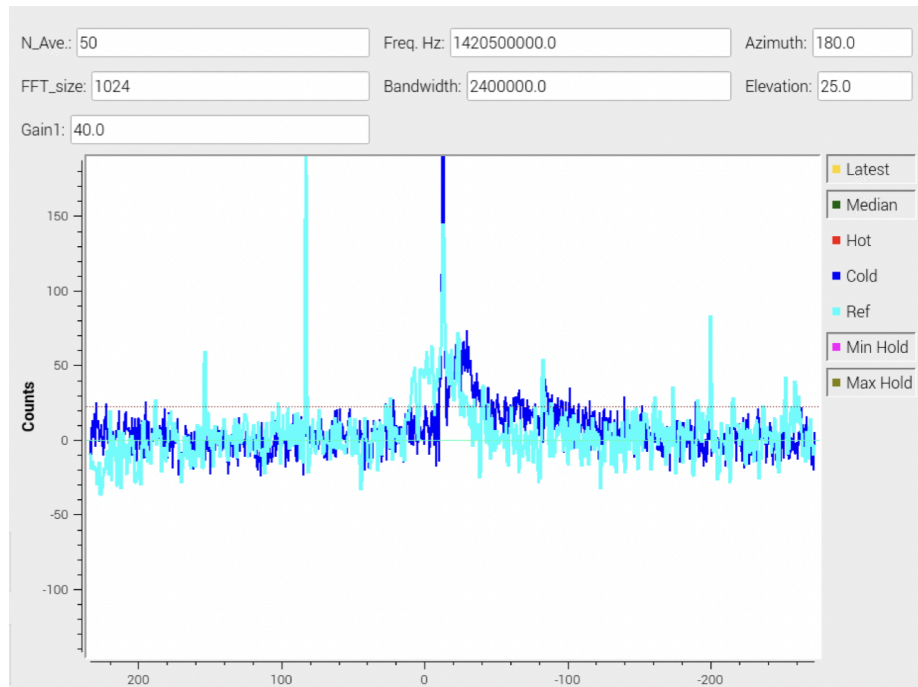

Figure 9: An example of what live data collection looks like. In this image the cold load is referenced from a different point in the sky that yields hydrogen.

R.py is used to process and visualize raw data from hydrogen line observations. The script handles various aspects of data processing, including the normalization of integration times, correction for frequency shifts, and mitigation of Radio Frequency Interference (RFI). Upon loading the raw observation data, the script applies frequency corrections relative to the known hydrogen line frequency (1420.40575 MHz) and normalizes the data to account for variations in integration time across different observations. It also flags and interpolates over known RFI features, ensuring that these artifacts do not skew the analysis. The script then plots the data using matplotlib. (Langston et al., 2017)

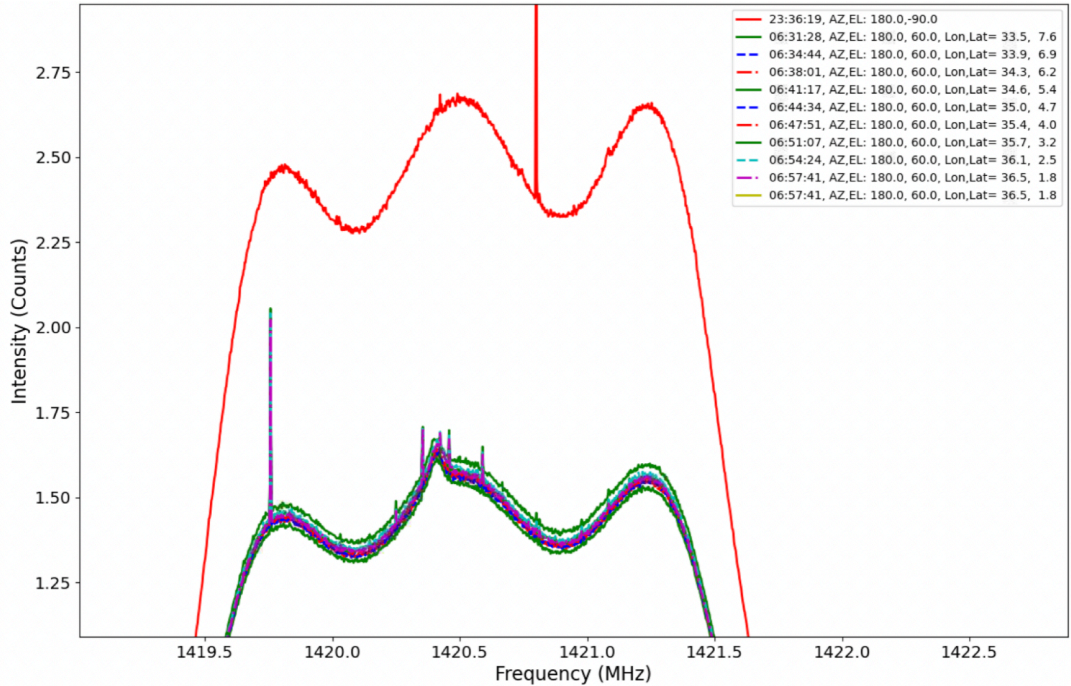

Figure 10: Example R plot. The red is the hot load while the lines below are the observations.

# Data analysis

T.py was mainly used for analysis as it allows for the extraction of meaningful scientific information from radio observations.

Data collection involves recording spectra using a radio telescope, typically with hot and cold load measurements for calibration. The cold load is the measurement of the telescope pointed at 90 degrees or any "dark" portion of the sky while the hot load is a result of the telescope at -90 degrees or facing the ground. These loads are used to calibrate the observational data by correcting for the system's gain and noise characteristics.

The raw observational data, as well as the hot and cold load data, are read into the processing pipeline. If specific hot and cold load files are provided, they are used directly; otherwise, these loads are computed from the observational data. The spectra from these loads are averaged over the specified time interval to improve the signal-to-noise ratio (SNR), which is ideal for detecting the weak hydrogen line signal against the background noise.

The processing pipeline also includes a flagging step where known RFI frequencies are identified, and the data is interpolated across these regions to mitigate the impact of interference. Additionally, median filtering is applied to the data to further reduce noise without compromising the integrity of the signal.

The observed spectra often contain a baseline component due to instrumental effects and residual atmospheric emission. This baseline must be removed to isolate the hydrogen line emission. The processing pipeline fits a polynomial baseline to the data, which is then subtracted from the observed spectrum. The order of the polynomial fit is adjustable based on the characteristics of the baseline.

After baseline subtraction, the script calculates the integrated intensity of the hydrogen line across a defined velocity range. This integration provides a measure of the total emission from neutral hydrogen in the observed region of the sky. The intensity-weighted velocity is also computed, giving the average velocity of the hydrogen gas along the line of sight.

The velocity axis is corrected for the Earth's motion relative to the barycenter of the solar system, ensuring that the observed velocities are accurate and can be compared with other observations.

The final step in the data processing pipeline involves preparing the calibrated spectra for scientific analysis. The processed spectra, now calibrated in physical units, are ready for further analysis, including the extraction of astrophysical parameters and comparison with theoretical models. (Langston, 2017)

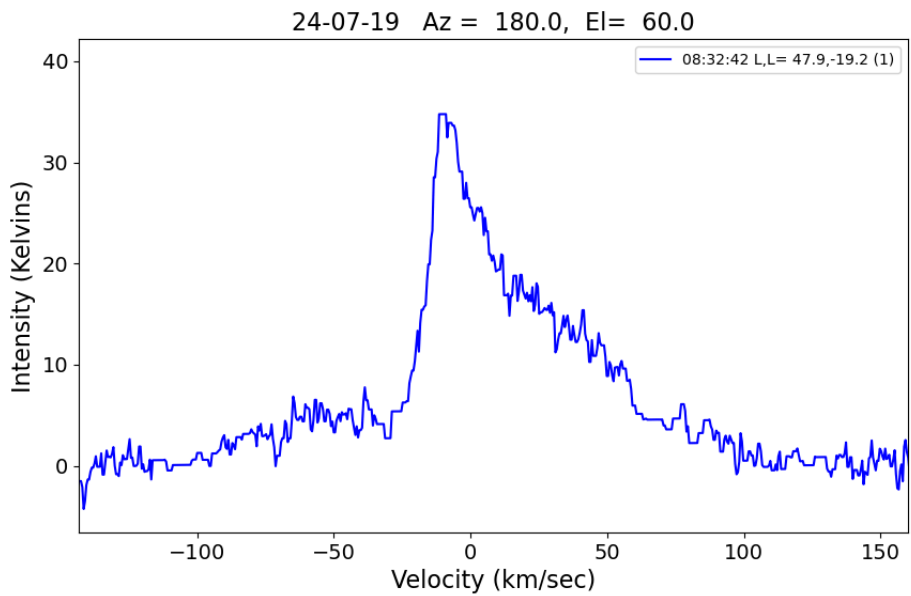

Figure 11: T.py example

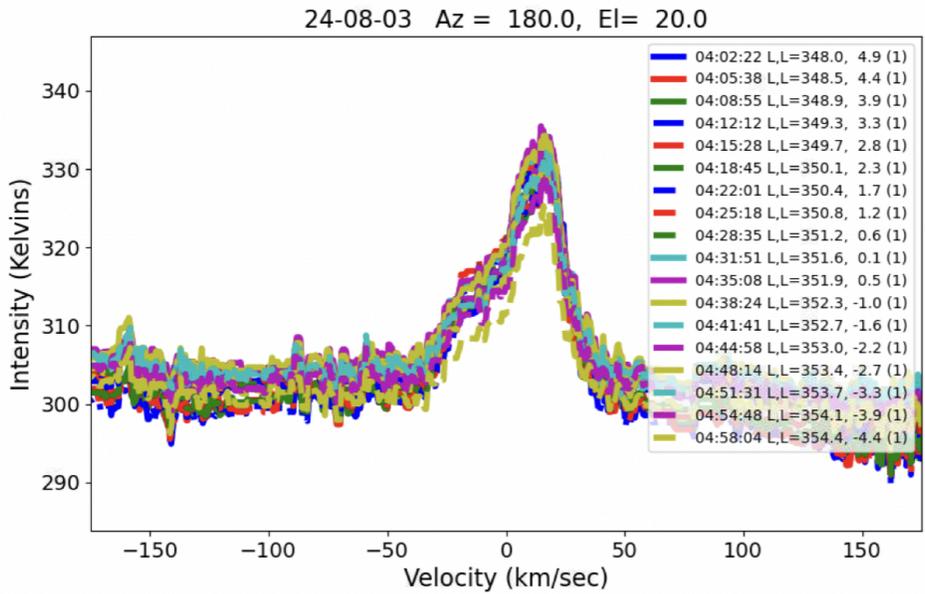

Figure 12: 1 hour observation at 180º south and 20º above the horizon.

This observation was conducted at an azimuth of 180.0° and an elevation of 20.0° over the course of 1 hour. The relatively low elevation could introduce atmospheric distortions, potentially affecting the accuracy of the data. The observation spanned galactic longitudes from 348.0° to 354.4°, corresponding to regions near the Galactic Center, particularly within the Sagittarius Arm, known for its dense molecular clouds and star-forming regions. The spectral data showed intensity values ranging from approximately 290 K to 340 K over a velocity range of -40km/s to +40km/s. It is important to note that the intensity of the data has not been baseline calibrated and the temperature of the line includes the relative temperature of the system. The peaks were slightly red shifted at around 25 Km/s, indicating relative motions of hydrogen clouds away from the observer. A secondary peak with less intensity is observed to be blue shifted at around -10 to -15 Km/s. These observed peaks reflect the dynamics within the hydrogen clouds, influenced by the Milky Way's rotation and localized phenomena. The consistency of these observations with the expected structure of the Sagittarius Arm supports the accuracy of the data, despite the potential atmospheric challenges.

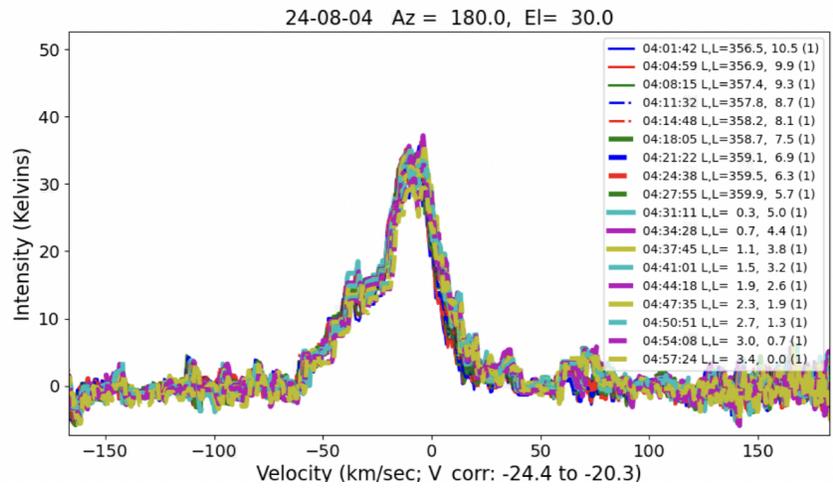

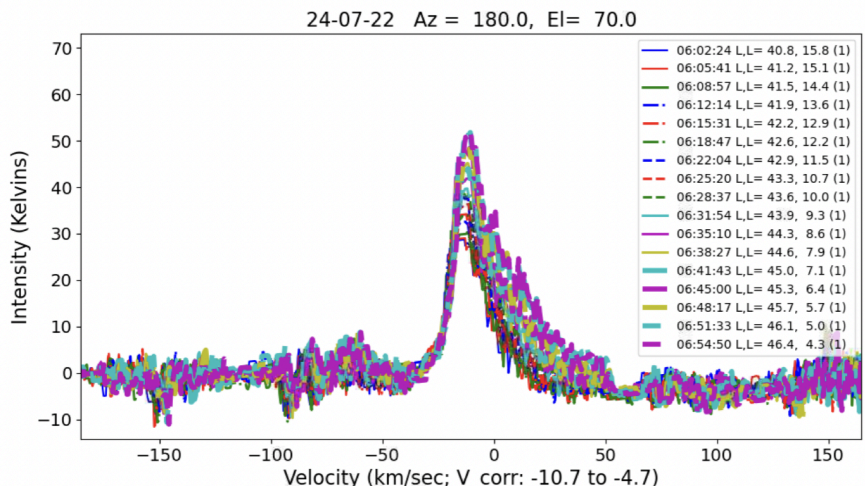

Figure 13: Single hour observations. Both observations occurred at 180º south. The first was at 30º above the horizon while the second was 70º above the horizon.

In the upper graph (24-08-04, Az = 180.0°, El = 30.0°), the primary peak is located at 0 km/s. The secondary peak at around -35 km/s indicates the presence of hydrogen gas moving towards the observer at a substantial relative velocity. This velocity shift is typically associated with regions of the Milky Way that lie closer to the galactic center, where the rotational velocity of gas is higher than in the Sun's local region. Specifically, the -35 km/s feature is consistent with hydrogen clouds in the Sagittarius-Carina Arm, one of the Milky Way's prominent inner spiral arms. The Sagittarius-Carina Arm is rich in neutral hydrogen and contains numerous star-forming regions, making it a major contributor to the detected hydrogen line emissions. The broad spread of velocities around this secondary peak, extending towards -50 km/s, indicates the presence of additional hydrogen clouds moving at different relative velocities, potentially from overlapping or adjacent regions within this arm. The relatively low elevation angle of 30.0° implies that the observation is being made closer to the galactic plane, where interstellar gas density is higher, contributing to the more complex velocity structure observed in the graph. The velocity correction of approximately -24.4 to -20.3 km/s suggests an adjustment for the Sun's motion relative to the observed gas, further aligning this emission with the Sagittarius-Carina Arm. The primary peak near 0 km/s represents gas moving in sync with the Sun's local standard of rest (LSR), possibly from the Local Arm (Orion Spur).

In the lower graph (24-07-22, Az = 180.0°, El = 70.0°), the main peak occurs around -25 km/s, indicating a strong hydrogen line signal from gas moving towards the observer at a moderate velocity. This peak likely corresponds to hydrogen clouds located in the Perseus Arm, a more distant spiral arm located outside the Sun's orbit around the Milky Way. The Perseus Arm is another major star-forming region that contains vast amounts of neutral hydrogen. The high elevation angle of 70.0° allows for a clearer observation with less atmospheric interference, resulting in a more defined peak and less noise compared to the upper graph. The velocity correction range of -10.7 to -4.7 km/s suggests a smaller adjustment for the Sun's motion, which aligns well with the typical velocities of hydrogen clouds in the outer spiral arms. The tail extending to positive velocities in the right side of the graph likely represents gas clouds moving away from the observer, possibly from outer regions of the Local Arm or inter-arm areas between the Perseus and Local arms. This broader velocity distribution suggests that the observer is capturing gas clouds at varying distances and rotational velocities, providing insight into the complex dynamics of the Milky Way's spiral structure. The combination of the main peak at -25 km/s and the positive velocity tail emphasizes the rotational motion of the galaxy, with gas in the outer regions lagging behind the Sun's orbit, thus appearing redshifted.

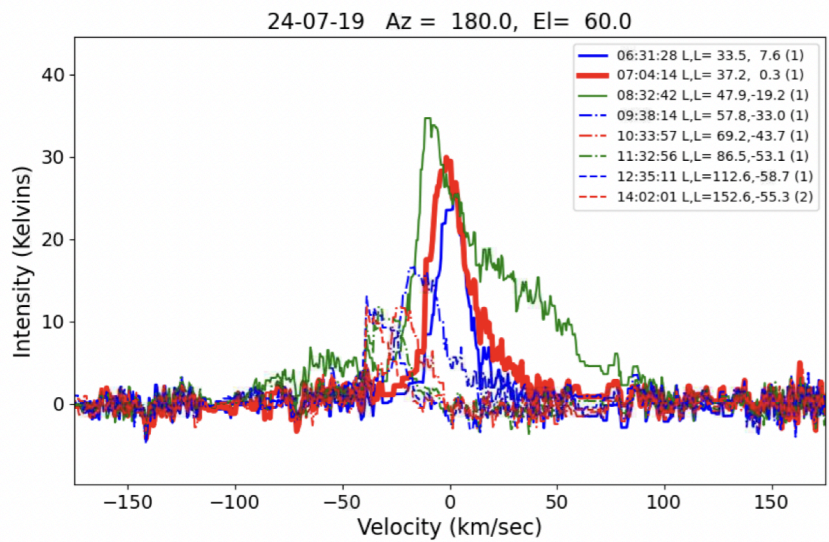

Figure 14: 8 hour observation at 180º south and 60º election above the horizon.

This observation was made at an azimuth of 180.0° and a higher elevation of 60.0°, reducing atmospheric interference. The observation covered a broader range of galactic longitudes, from 33.5° to 152.6°, spanning multiple spiral arms, including the Sagittarius, Scutum-Centaurus, and Perseus Arms. The observational period lasted about 8 hours. This observation was calibrated so that the baseline is at 0K and the intensity depicted are the temperatures of just the Hydrogen line emissions. The intensity ranged from 0 K to approximately 40 K, with spectral peaks distributed across various velocities, which is expected since multiple arms were observed during the observation. The variation in intensity and peak positions reflects the transition from denser regions near the Galactic Center to more peripheral areas of the galaxy. These observations are consistent with the expected distribution of neutral hydrogen across different spiral arms of the Milky Way.

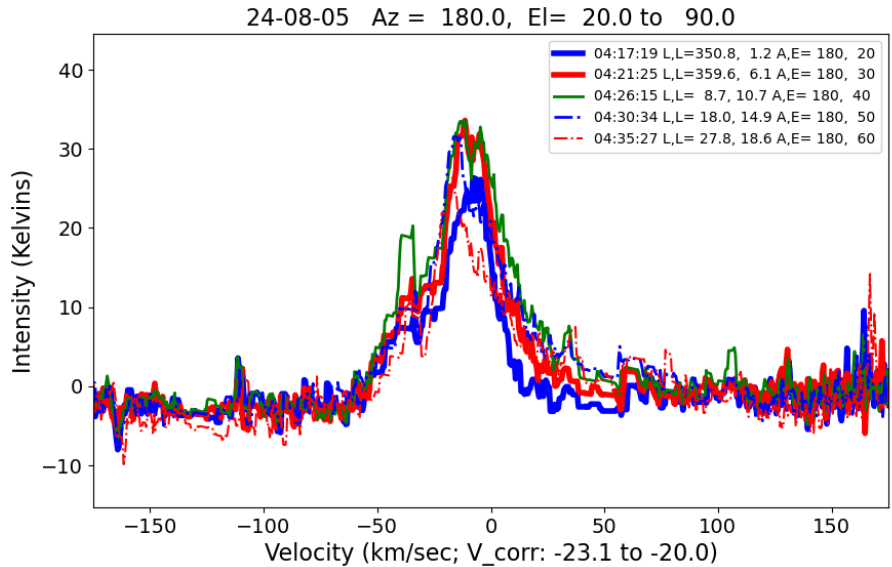

Figure 15: Varying Elevations at 180º south

This graph displays hydrogen line emissions observed at different elevations (El = 20° to 90°) along a fixed azimuth (Az = 180.0°), revealing the motion of hydrogen gas in various parts of the Milky Way. The central peak near 0 km/s corresponds to gas moving with the observer's local standard of rest, likely from the Local Arm (Orion Spur), and becomes sharper and more intense at higher elevations (El = 90°), indicating reduced atmospheric interference. On the negative velocity side, a blueshifted feature between -100 km/s and -150 km/s is visible at all elevations, likely due to gas in the Sagittarius-Carina Arm moving toward the observer. The positive velocity side shows a broad redshifted tail (50 to 150 km/s), representing gas receding from the observer, possibly from the Perseus Arm. Higher elevations provide clearer signals, while lower elevations show more atmospheric noise, but the overall structure of the Milky Way's spiral arms is still detectable at all angles.

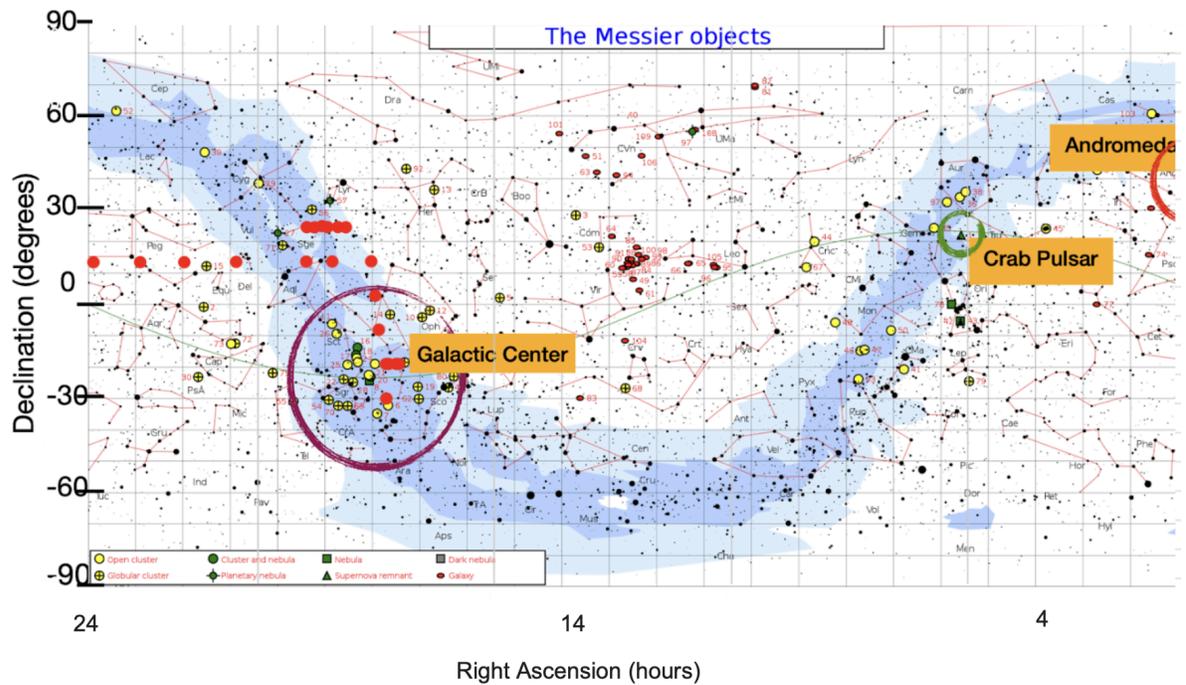

Figure 16: Locations of observations. Background image (Cornmell)

The red dots representing the observations are spread across various regions of the sky, primarily concentrated along the plane of the Milky Way, which is marked by the blue band on the map. A notable concentration of these dots is in the region surrounding the Galactic Center, located near the constellations Sagittarius and Scorpius.

In relation to the earlier line analyses, the hydrogen peaks seen around -35 km/s and -25 km/s in the graphs are consistent with observations in the Sagittarius-Carina Arm and regions close to the Galactic Center. These red dots near the Galactic Center reinforce the observations and confirm that the setup is picking up hydrogen gas clouds from this arm, as well as possibly from the Scutum-Centaurus Arm, which is also visible in this area.

Additionally, the red dots extending along the plane of the Milky Way towards Cygnus and Aquila correspond well with the analyses of hydrogen gas clouds in the outer regions of the Milky Way, particularly those associated with the Perseus Arm. The wide spread of red dots along the galactic plane aligns with the fact that the hydrogen gas is distributed throughout the Milky Way's spiral arms, and the observations are capturing these key regions. Overall, the concentration of red dots, especially near the Galactic Center and along the Milky Way, confirms that the observations are aligned with regions rich in hydrogen gas and star-forming activity, consistent with the peaks observed in the velocity profiles.

# Acknowledgments

I would like to thank Dr. Glen Langston at the NSF for the documentation he has created that enables students like myself to undertake projects like this. I would also like to thank him for providing me with software help, taking the time to meet with me, and inspiring me.

I also want to thank Ava Polzin at the University of Chicago for her continued support throughout this entire project and for advising me in the areas of cosmology, instrumentation, general astrophysics and data presentation.

I would also like to thank SARA, the RTL-SDR community, and the authors of all lightwork memos associated with WVURAIL for guiding me and inspiring me.

I want to extend my gratitude to my computer science teacher Shoshana Arunasalam who has supported me with the programming and data visualization aspects of the project.